\documentclass[twocolumn,nofootinbib,prl,preprintnumbers,amsmath,amssymb]{revtex4-1}
\usepackage{natbib,hyperref}
\usepackage{amsmath}
\usepackage{graphicx}
\usepackage{dcolumn}
\usepackage{bm}
\usepackage{epsfig,color,xspace,multirow,xr,bbold}
\usepackage[all]{xy}
\usepackage{setspace}
\usepackage{url}
\usepackage[colorinlistoftodos]{todonotes}
\usepackage{threeparttable} 
\newcommand{\Ref}[1]{Ref.~\onlinecite{#1}}

\usepackage{natbib,hyperref}

\begin{document}

\title{
Exact separation of radial and angular correlation energies in two-electron atoms
}
\date{\today}     
\author{Anjana R Kammath$^{1}$}
\affiliation{$^1$Indian Institute of Science Education and Research, Mohali, Punjab 140306, India}
 \affiliation{$^2$VSRP fellow at Tata Institute of Fundamental Research, Centre for Interdisciplinary Sciences, Hyderabad 500107, India}

\author{Raghunathan Ramakrishnan$^{2}$}
\email{ramakrishnan@tifrh.res.in}
 \affiliation{$^2$Tata Institute of Fundamental Research, Centre for Interdisciplinary Sciences, Hyderabad 500107, India} 


\begin{abstract}
Partitioning of helium atom's correlation energy into radial and angular contributions, although of fundamental interest, has eluded critical scrutiny.  Conventionally, radial and angular correlation energies of helium atom are defined for its ground state as deviations, from Hartree--Fock and exact values, of the energy obtained using a purely radial wavefunction devoid of any explicit dependence on the interelectronic distance.  Here, we show this rationale to  associate the contribution from radial-angular coupling entirely to the angular part underestimating the radial one, thereby also incorrectly predict non-vanishing residual radial probability densities. We derive analytic matrix elements for the high-precision Hylleraas basis set framework to seamlessly uncouple the angular correlation energy from its radial counterpart. The resulting formula agrees with numerical cubature yielding precise purely angular correlation energies for the ground as well as excited states. Our calculations indicate 60.2\% of helium's correlation energy to arise from strictly radial interactions; when excluding the contribution from the radial-angular coupling, this value drops to 41.3\%. 
\end{abstract}

\maketitle
High-precision variational calculations of two-electron atoms have rigorously enabled quantitative agreement of first-principles predictions with such subtle physical measurements as relativistic and Lamb shift contributions to atomic transition frequencies~\cite{yan1995high}. Arguably, the most critical prerequisite for reaching such accuracies is an explicit dependence of variational trial functions on the interelectronic separation, $u=|{\bf r}_1 - {\bf r}_2|$. Historically, it was the inclusion of this variable along with $s=r_1+r_2$ and $t=r_2-r_1$, in the wavefunction that enabled Hylleraas to predict the ground and first excited states of helium very accurately~\cite{hylleraas1964schrodinger}. 

Correlation energy of helium is the difference between its non-relativistic exact ground state energy, $E^{\rm exact}_0=-2.903724377034119598311$~au~\cite{nakashima2007solving}, and its Hartree--Fock (HF) limit, $E^{\rm HF}_0=-2.861679995612$~au~\cite{raffenetti1973even}, i.e., $E_{\rm c} = E^{\rm exact}_0 - E^{\rm HF}_0=-0.0420443814221194$~au. Taylor and Parr~\cite{taylor1952superposition} conjectured that a complete wavefunction in $r_1$  and $r_2$ should account for only that part of correlation energy arising from the motion of the electrons radially towards and away from the nucleus. Using a modified configuration-interaction approach, employing basis functions that depend on ordered radial coordinates ($r_<$ and $r_>$), Goldman has established the radial limit of the ground state energy of helium very precisely as $-2.879028767319214408538$~au~\cite{goldman1997accurate}. This value when used along with the HF limit yields $-0.017348771707$ au as the atom's radial limit of the correlation energy, which amounts to only 41.3\% of $E_c$. The remaining 58.7\% of $E_c$ that is captured only when the two-electron wavefunction is explicitly made a function of $u$ is conventionally defined as the angular correlation energy~\cite{lennard1952liii,moiseyev1975coupling,wilson2010electron,saha2003radial}---because via $u$ enters the third independent coordinate $\theta$, the angle subtended between the two position vectors ${\bf r}_1$ and ${\bf r}_2$. To date, insufficient efforts have gone into critically inspecting the validity of such an additive interpretation of $E_c$. In the 50s, Green and others~\cite{green1953correlation,green1954correlation} have employed few-parameters Hylleraas~\cite{hylleraas1964schrodinger} and Chandrasekhar~\cite{chandrasekhar1953shift} wavefunctions in configuration-interaction calculations to partition $E_0$ into $E^{\rm HF}_0$, and three correlation terms of radial, angular and mixed characters. This approach, however, resulted in $E_c=-0.0688$ au for helium, largely overestimating the actual value. Later, the calculations of Moiseyev revealed the coupling between radial and angular components of $E_c$ to emerge in high order terms of $1/Z$-perturbation theory~\cite{moiseyev1975coupling}.

The purposes of this letter are to, firstly, expose the residual radial correlation that is present in an exact wavefunction and is not captured by a fully radial wavefunction devoid of $u$-dependence. Then, we  present a new strategy based on analytic matrix elements to compute precise pure angular correlation energies, $E_c^{\rm ang}$, for helium and its isoelectronic ions. We formulate the problem using the explicitly correlated wavefunction {\emph Ansatz} \begin{eqnarray}
\Psi \left( {\bf r}_1, {\bf r}_2 \right) = e^{ -\alpha s/2 }  
\sum_{l,m,n=0}^{l+m+n\le \Omega} C_{lmn}s^{l^\prime} t^{m^\prime} u^{n^\prime} 
\end{eqnarray}
where $l,m,n\in {\bf Z}$. The convergence of the trial function is studied by progressively increasing the number of basis functions, which varies as $N =(\Omega+1)(\Omega+2)(\Omega+3)/6$~\cite{drake2002ground}. We have optimized the exponent $\alpha$ in all our calculations performed with quadruple precision. For the choice of $l^\prime=l-n$,  $m^\prime=2m$ and $n^\prime=n-2m$, we obtain a Kinoshita wavefunction ~\cite{kinoshita1957ground}, which for $\Omega=11$ ($N=364$) results in the exact ground state energy $-2.90372438$ au. Alas, the Kinoshita wavefunction converges rather poorly for the excites states. So, in this study, we have computed the energies of $1s^2$, $1s^12s^1$, and $1s^13s^1$  states of Helium using the Hylleraas formalism, employing $l^\prime=l$,  $m^\prime=2m$ and $n^\prime=n$, as $-2.90372438$ au ($\Omega=12$), $-2.14597405$ au ($\Omega=12$), and $-2.06127200$ au ($\Omega=20$). These values agree, to the reported precision, with those from the double-basis set variational calculations of Drake and Yan~\cite{drake1994variational}. 



%

Purely radial Hylleraas wavefunctions are obtained by setting $n^\prime=0$,
satisfying $N =(\Omega+1)(\Omega+2)/2$.
\begin{eqnarray}
\Psi^{\rm rad} \left( {\bf r}_1, {\bf r}_2 \right) = e^{ -\alpha s/2 }  \sum_{l,m=0}^{l+m\le \Omega} 
C_{lm}s^{l^\prime} t^{m^\prime} 
\end{eqnarray}
With $\Omega=40$, along with variationally optimized $\alpha$, we 
obtain the energies of $1s^2$, $1s^12s^1$, and $1s^13s^1$  states of Helium as  
$-2.87902846$,    $-2.14419704$, and     
$-2.06079381$ au, respectively, the ground state energy deviating from 
Goldman's precise value~\cite{goldman1997accurate} by merely $3\times10^{-7}$ au.   
Koga had earlier noted superior convergence of the ground state energy using a radial
Kinoshita wavefunction with $l^\prime=l-m$, and $m^\prime=2m$~\cite{koga1996radial}.
 With these constraints we obtain
the improved values, $-2.87902875$, $-2.14419727$, and $-2.06079404$~au, 
for the lowest three singlet states of helium. By modifying the radial Kinoshita 
framework as an optimal $N-$term wavefunction, as proposed by Koga~\cite{koga1996radial},
\begin{eqnarray}
\Psi^{\rm rad}\left( {\bf r}_1, {\bf r}_2 \right) = e^{ -\alpha s/2 } \sum_{i=0}^{N} C_{i} s^{l^\prime} t^{m^\prime} 
\end{eqnarray}
we find a more precise radial limit of helium's ground state energy 
converging to $-2.8790287673153$ au for $N=43$. In this case, we have varied $m^\prime=m$
as a positive integer, and ensured one of the $N$ terms to be 
of singlet-spin type with $l=0$, and $m=0$.

\begin{figure}[hpt] 
\centering  
\includegraphics[width=8.8cm, angle=0.0]{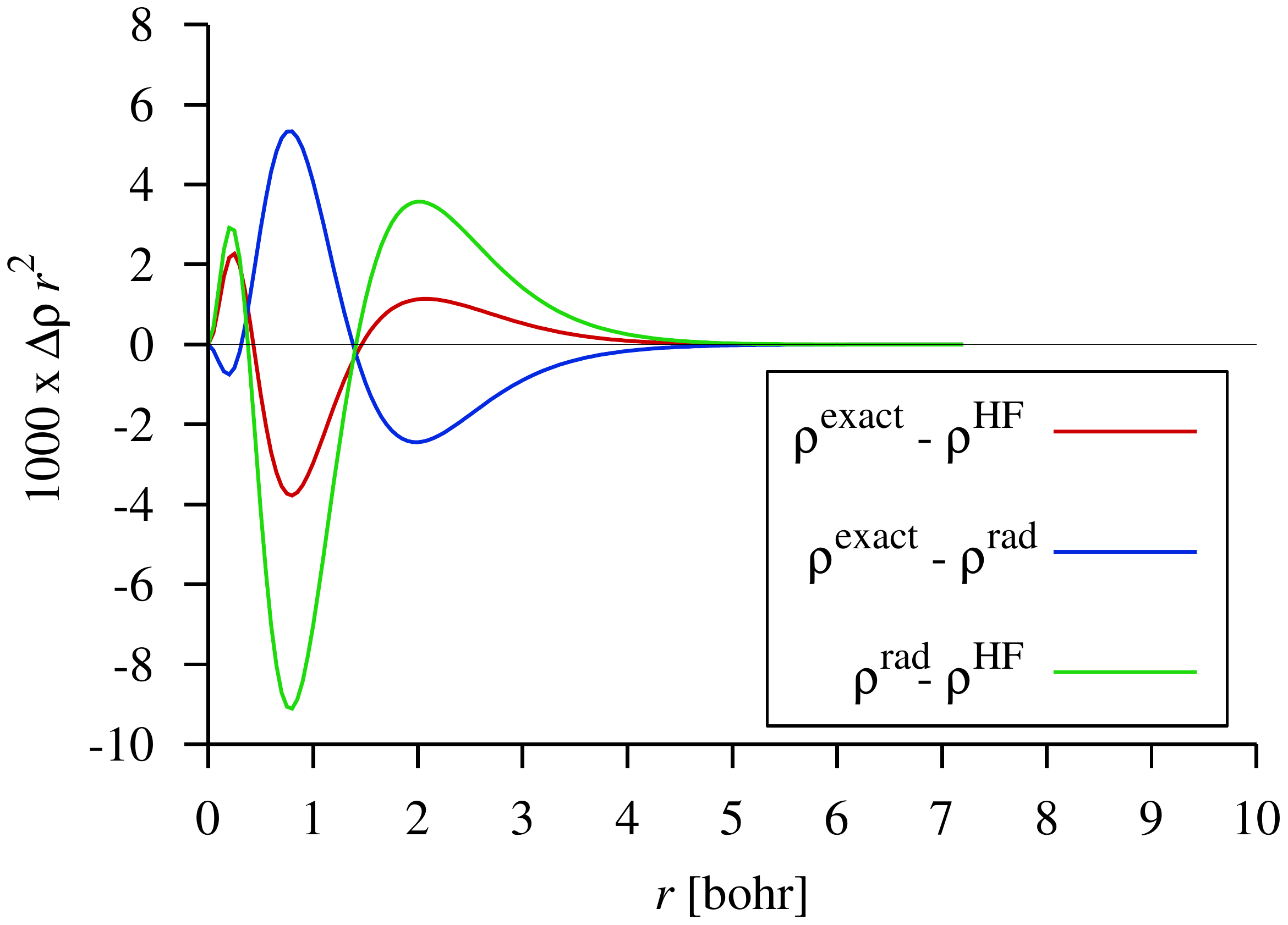}
\caption{Differences between exact, fully radial, and uncorrelated (i.e. HF) radial probability densities of helium. HF wavefunction
was computed using six Slater-type functions with optimal exponents.}
\label{fig:probdens1}
\end{figure}

With such a precise energy estimation, the corresponding
radial wavefunction is expected to
capture all the radial dependence beyond that of HF. 
To further elucidate the point, let us now zero-in on the reduced radial density function 
\begin{eqnarray}
\rho\left( \vec{r}_1\right) & = & 
8 \pi^2 \int_{0}^{\infty} r^2_2dr_2
\int_{0}^{\pi}\sin\theta d\theta \, 
|\Phi_{i} \left( {\bf r}_1, {\bf r}_2 \right) |^2
\end{eqnarray}
where $\Phi$ can be 
$\Psi\left( {\bf r}_1, {\bf r}_2 \right)$, $\Psi^{\rm rad}\left( {\bf r}_1, {\bf r}_2 \right)$ 
or $\Psi^{\rm HF}\left( {\bf r}_1, {\bf r}_2 \right)$. For all three
wavefunctions, the reduced density function follows 
$\int_{0}^{\infty}\rho\left( \vec{r}_1\right)r^2_1dr_1 =1$. 

In Fig.~\ref{fig:probdens1}, we find the change in $\rho$ while 
going from the HF to an exact wavefunction to be different than while going from an 
HF wavefunction to an exclusively radial one. Such a trend implies the exact wavefunction to capture radial 
correlation that is coupled to the angular degree of freedom and is inaccessible to $\Psi^{\rm rad}$ lacking $u$-dependence.  
While purely radial correlation has the effect of larger divergence in density from that of HF, 
the radial correlation coupled with the angular counterpart has the opposing effect of 
bringing the electron density closer to the HF one. 
Subtracting $\rho^{\rm rad}$ from the 
exact $\rho$ indeed reveals such a trend (Fig.~\ref{fig:probdens1}). 

\begin{figure}[hpt] 
\centering  
\includegraphics[width=8.8cm, angle=0.0]{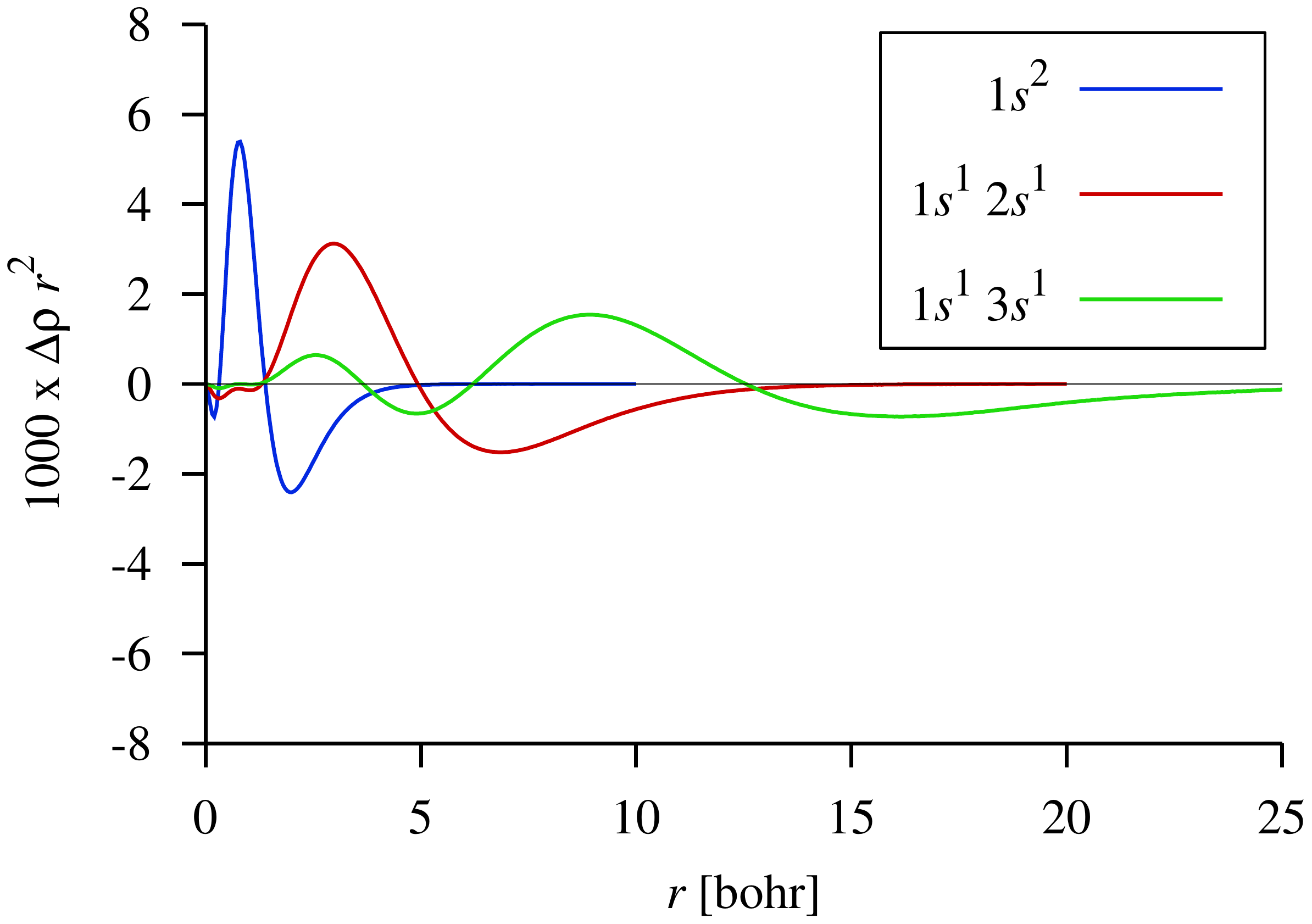}
\caption{Differential radial probability density 
$\Delta \rho=\rho^{\rm exact}-\rho^{\rm rad}$
showing residual correlation for the lowest three singlet $S$-states of He.}
\label{fig:probdens2} 
\end{figure}

The situation is similar also in the case of the
$1s^12s^1$ and $1s^13s^1$ excited states. For both, we find the probability density from a purely 
radial wavefunction to deviate from that of an exact wavefunction~(see Fig.~\ref{fig:probdens2}). 
However, for the excited states, the residual radial correlation seems to decrease with increase in energy. At this point, it is worth noting that---as pointed out in \Ref{moiseyev1975coupling}---the total correlation in excited states is essentially angular. Later we will quantify $E_c^{\rm ang}$ for these states precisely.

\begin{table*}[htp!]
\begin{threeparttable}[b]  
\caption{Convergence in the analytic values of angular kinetic energies, 
$T^{\rm ang}_{\rm ana}$,
with number of terms, $N_t$, in Eq.~11. $\Omega $ determines the truncation of the
Hylleraas wavefunction (see Eq.~1).
Results from numerical cubature\cite{hahn2005cuba} calculations are listed for comparison, $T^{\rm ang}_{\rm num}$. 
All values are in au.}
        \begin{tabular}{r r r r r r r r } 
        \hline
\multicolumn{1}{l}{$\Omega$}  & 
\multicolumn{6}{l}{~~~$T^{\rm ang}_{\rm ana}$} & 
\multicolumn{1}{l}{~~~$T^{\rm ang}_{\rm num}$}\\
\cline{2-7}
\color{black}{}  & 
\multicolumn{1}{l}{~~~$N_t$=1} & 
\multicolumn{1}{l}{~~~$N_t$=5} & 
\multicolumn{1}{l}{~~~$N_t$=25} & 
\multicolumn{1}{l}{~~~$N_t$=125} & 
\multicolumn{1}{l}{~~~$N_t$=625} & 
\multicolumn{1}{l}{~~~$N_t$=3125} & 
\color{black}{}\\
\hline
     1&   ~~~0.01523418   &~~~0.01711083 &  ~~~0.01741473  & ~~~0.01742332    &~~~0.01742342  & ~~~0.01742342&~~~0.01742342\\
     2&      0.01543591   &   0.01696821 &     0.01716376  &    0.01716805    &  0.01716809  &    0.01716809&~~~0.01716809\\
     4&      0.01519800   &   0.01661683 &     0.01678819  &    0.01679176    &  0.01679179  &    0.01679179&~~~0.01679179\\
     8&      0.01516327   &   0.01657599 &     0.01674755  &    0.01675117    &  0.01675121  &    0.01675121&~~~0.01675121\\
    16&      0.01516252   &   0.01657524 &     0.01674683  &    0.01675046    &  0.01675049  &    0.01675049&~~~0.01675049\\

\hline
\end{tabular}
 \label{tab:tang}

\end{threeparttable}  
\end{table*}

We now divert our attention to the exact separation of radial and angular correlation
energies based on analytic expressions for the matrix elements. 
Our derivation is grounded on the fact that 
$\Psi^{\rm HF}\left( {\bf r}_1, {\bf r}_2 \right)$ is the variationally best radial wavefunction separable in ${\bf r}_1$ and ${\bf r}_2$, lacking any dependence on $u$. Hence,
separating the kinetic energy terms that are dependent on the $u$ variable should provide 
angular correlation energy via the virial theorem $E_c^{\rm ang}=-T^{\rm ang}$. 
We begin our derivation with the kinetic energy operator in the 
$s$, $t$ and $u$ variables
\begin{eqnarray}
\hat{T}_{stu}&=&
 -\left( \partial^2_s + 
\partial^2_t +\partial^2_u  \right)  - \frac{4s}{ \left(s^2-t^2 \right)} \partial_s +  \frac{4t}{ \left(s^2-t^2 \right)}\partial_t   \nonumber \\
&&- \frac{2 }{u } \partial_u
- \frac{2 s \left(u^2-t^2 \right)}{u \left(s^2-t^2 \right)} \partial^2_{s,u}
- \frac{2 t \left(s^2-u^2 \right)}{u \left(s^2-t^2 \right)} \partial^2_{t,u} 
\end{eqnarray} 
For our purpose, it is vital to decouple the kinetic energy operator 
as $\hat{T}_{stu}=\hat{T}_{stu}^{\rm rad}+\hat{T}_{stu}^{\rm ang}$;
individual terms defined as
\begin{eqnarray}
\hat{T}_{stu}^{\rm rad}=-\frac {1}{2 }\sum_{i=1}^2r_i^{-2}\partial_{r_i}r_i^{2}\partial_{r_i}; \quad 
\hat{T}_{stu}^{\rm ang}=-\frac {1}{2 }
\nabla_{\theta}^2, 
\end{eqnarray} 
but retain the Hylleraas' coordinates representation that facilitates analytic computation of the matrix elements. To this end, we invoke substitutions
$r_1=(s-t)/2$, $r_2=(s+t)/2$, and $u=r_{12}$. With the latter quantity defined as
\begin{center}
$u^2=r_{12}^2= r_1^2 +r_2^2 - 2r_1r_2\cos\theta$
\end{center}
we arrive at $udu = r_1r_2\sin\theta d\theta$. 
A purely angular Laplacian can now be written as
\begin{eqnarray}
\nabla_{\theta}^2 &=& \left[r_1^{-2}(\sin \theta)^{-1}  +
r_2^{-2}(\sin \theta)^{-1}\right]\partial_\theta\left(\sin \theta \partial_\theta\right).
\end{eqnarray}
Direct substitutions of $r_1,r_2,\theta$ along with 
$\sin^2\theta=1-\cos^2\theta $ results in a more useful
expression which is directly expressed in the Hylleraas coordinates as
\begin{eqnarray}
\nabla_{\theta}^2 &=& \left[\frac{1}{u(s-t)^2}+\frac{1}{u(s+t)^2}\right] \nonumber \\
& & \partial_u \left\lbrace\frac{-u^4-s^2t^2+(s^2+t^2)u^2}{u}\right\rbrace\partial_u
\end{eqnarray}
This Laplacian  when operating on a primitive basis function 
$|l^\prime,m^\prime,n^\prime \rangle =e^{ -\alpha s/2 } s^{l^\prime} t^{m^\prime} u^{n^\prime}$ yields
\begin{eqnarray}
\nabla_{\theta}^2 |l^\prime,m^\prime,n^\prime \rangle  &=&|l^\prime,m^\prime,n^\prime \rangle  \left[\frac{n^\prime(s-t)}{(s+t)}+\frac{n^\prime(s+t)}{(s-t)}\right]  \\ 
& &  n^\prime(s^2+t^2)u^{-1}-(n^\prime+2)u-(n^\prime-2)s^2t^2u^{-3} \nonumber
\end{eqnarray}
where we have multiplied the resulting expression with the volume element $u(s^2-t^2)$.

Deriving the angular kinetic energy matrix elements is now readily accomplished 
with the use of Maclaurin series for $1/\left( s-t\right)$ and 
$1/\left( s+t\right)$
yielding
\begin{eqnarray}
&& \langle l_i^\prime m_i^\prime n_i^\prime |\nabla_{\theta}^2|l_j^\prime m_j^\prime n_j^\prime\rangle = \nonumber \\
& & \int_0^\infty ds \int_0^s du \int_0^u dt e^{-\alpha s} s^{L} t^{M} u^{N} n_j^\prime s^{-1}  \nonumber \\ 
& & \left[n_j^\prime(s^2+t^2)u^{-1}-(n_j^\prime+2)u-(n_j^\prime-2)s^2t^2u^{-3}\right]  \nonumber \\
& & \left[(s+t)\sum_{k=0}^\infty \left(t/s\right)^k  + (s-t)\sum_{k=0}^\infty (-1)^k\left(t/s\right)^k \right]
\end{eqnarray}
where $L=l_i^\prime+l_j^\prime$, $M=m_i^\prime+m_j^\prime$ and $N=n_i^\prime+n_j^\prime$.

Closed form expression for the angular kinetic energy matrix elements can then be written as a sum of two
series: one over odd indices and the other over even indices. 
\begin{widetext}
\begin{eqnarray}
\langle l_i^\prime m_i^\prime n_i^\prime |\hat{T}_{ang}| l_j^\prime m_j^\prime n_j^\prime \rangle & = &
-n_j^\prime
\sum_{k=0,2,\cdots}^{N_t} \left[
n_j^\prime{\mathcal I}(L+2-k,M+k,N-1) +
n_j^\prime{\mathcal I}(L- k,M+2+k,N-1) \right. -\nonumber\\
&& \left. (n_j^\prime+2){\mathcal I}(L-k,M+k,N+1) -
(n_j^\prime-2){\mathcal I}(L+2-k,M+ 2+k,N-3) \right] \nonumber\\
& & 
-n_j^\prime\sum_{k=1,3,\cdots}^{N_t} \left[ n_j^\prime{\mathcal I}(L+1-k,M+1+k,N-1) +
n_j^\prime{\mathcal I}(L-1-k,M+3+k,N-1) - \right. \nonumber\\
&& \left. (n_j^\prime+2){\mathcal I}(L-1 -k,M+1+k,N+1)- 
(n_j^\prime-2){\mathcal I}(L+1-k,M+3+k,N-3) 
\right]
\label{eq:Tang}
\end{eqnarray}
\end{widetext}
In the above equation, the primitive integral takes the usual form \cite{bethe2012quantum}
\begin{eqnarray}
{\mathcal I}\left(a,b,c\right) & = & \int_0^\infty ds \int_0^s du \int_0^u dt e^{-s} s^a t^b u^c \nonumber \\
& = & \frac{\Gamma \left( a+b+c+3 \right)}{\left( b+1 \right)\left( b+c+2 \right)}
\end{eqnarray}
We note in passing that the dependence on the exponent can be incorporated by the scaling relation
$T^{\rm ang}=\alpha^2 \langle \Psi |\hat{T}^{\rm ang} | \Psi \rangle /   \langle \Psi | \Psi \rangle$, 
where the factor $1/8$ in the volume element cancels out. To evaluate the accuracy of Eq.~\ref{eq:Tang},
we have performed calculations with Hamiltonian matrix elements  
computed using numerical cubature~\cite{hahn2005cuba} instead of analytic formulae. 
For a given $\alpha$, the results of these calculations agree perfectly with those
computed using analytic matrix elements. 
In Table \ref{tab:tang}, we compare selectively the matrix elements of 
the angular kinetic energy from both procedures. 
For various values of $\Omega$, we report the expectation value 
of $\hat{T}_{\rm ang}$ in the ground state. For $N_t=625$, we reach 
convergence in the series agreeing with cubature. The resulting value of 
$E_c^{\rm ang}=-T^{\rm ang}=-0.01675049$~au accounts for 39.8\% of total $E_c$. 
In contrast, difference between the ground state energies of $\Psi^{\rm exact}$ and
$\Psi^{\rm rad}$, as yet defined~\cite{wilson2010electron} as 
the angular correlation energy, is as high as 58.7\%.
The exact value of radial correlation energy can now be deduced as 
$E_c^{\rm rad}=E_c-E_c^{\rm ang}=-0.02529389$~au, 
and can be correctly identified as the dominant contributor to 
the total correlation energy of He. Hence we feel that the previous limit of 
$E_c^{\rm rad}$, defined as the difference between the energy obtained using a purely 
radial wavefunction and the HF energy, can at best be denoted as $E_c-E_c^{\rm ang}-E_c^{\rm ang-rad}/2$. 
\begin{table}[htp!]
\caption{Total and correlation energies of helium
for the lowest three $S$-states. Also given are 
average values of $s$, $t$ and $u$ along with their standard deviations. All values are in au.
}
        \begin{tabular}{l r r r } 
        \hline
\multicolumn{1}{l}{Property}&\multicolumn{1}{l}{~~~$1s^2$}&\multicolumn{1}{l}{~~~$1s^12s^1$}&\multicolumn{1}{l}{~~~$1s^13s^1$}\\
\hline
$E$ &~~~-2.90372438     &~~~-2.14597405&~~~-2.06127200\\
 $E^{\rm ang}_c$     &-0.01675049 & -0.00113042& -0.00030431\\
 $E^{\rm rad}_c$     &-0.02529389&            &           \\
 $\langle s\rangle$  &1.85894459 &5.94612193&13.02334671 \\
 $\langle t\rangle$  &0.65422575 &4.44779651&11.52344922 \\
 $\langle u\rangle$  &1.42207026 &5.26969586&12.30451548 \\
 $\sigma_s$          &0.76518757 &2.15794814 &4.53801299 \\
 $\sigma_t$          &0.55202685 &2.13562968 &4.53528083 \\
 $\sigma_u$          &0.70296195 &2.12900831 &4.52075484 \\
\hline
\end{tabular}
 \label{tab:tab2}
\end{table}

Our approach also enables the calculation of  $E_c^{\rm ang}$ for excited states as expectation values. However, to achieve precise results, the exponent $\alpha$ needs to be optimized for each state separately. The resulting values are collected along with the expectation values and standard deviations of $s$, $t$ and $u$ in Table~\ref{tab:tab2}.  The latter values are in close agreement with results from previous multi-configuration HF calculations~\cite{koga2010electron,koga2011interelectronic}.  The magnitudes of the standard deviations of these variables are of the same order as their respective average values indicating a broad spread of the wavefunctions in these variables.  Alas, it is not possible to determine $E_c^{\rm rad}$ for the $1s^12s^1$, and $1s^13s^1$ states because excited states in the HF theory are ambiguous. For instance, a previous study~\cite{ramakrishnan2012control} had shown the excited states within this model to be non-orthogonal to the ground state rendering linear superpositions impossible.  In Table~\ref{tab:tab2}, overall one notes  $E_c^{\rm ang}$ to gradually vanish with increasing energy and average inter-electronic distance.  

\begin{figure}[hpt] 
\centering  
\includegraphics[width=8.5cm, angle=0.0]{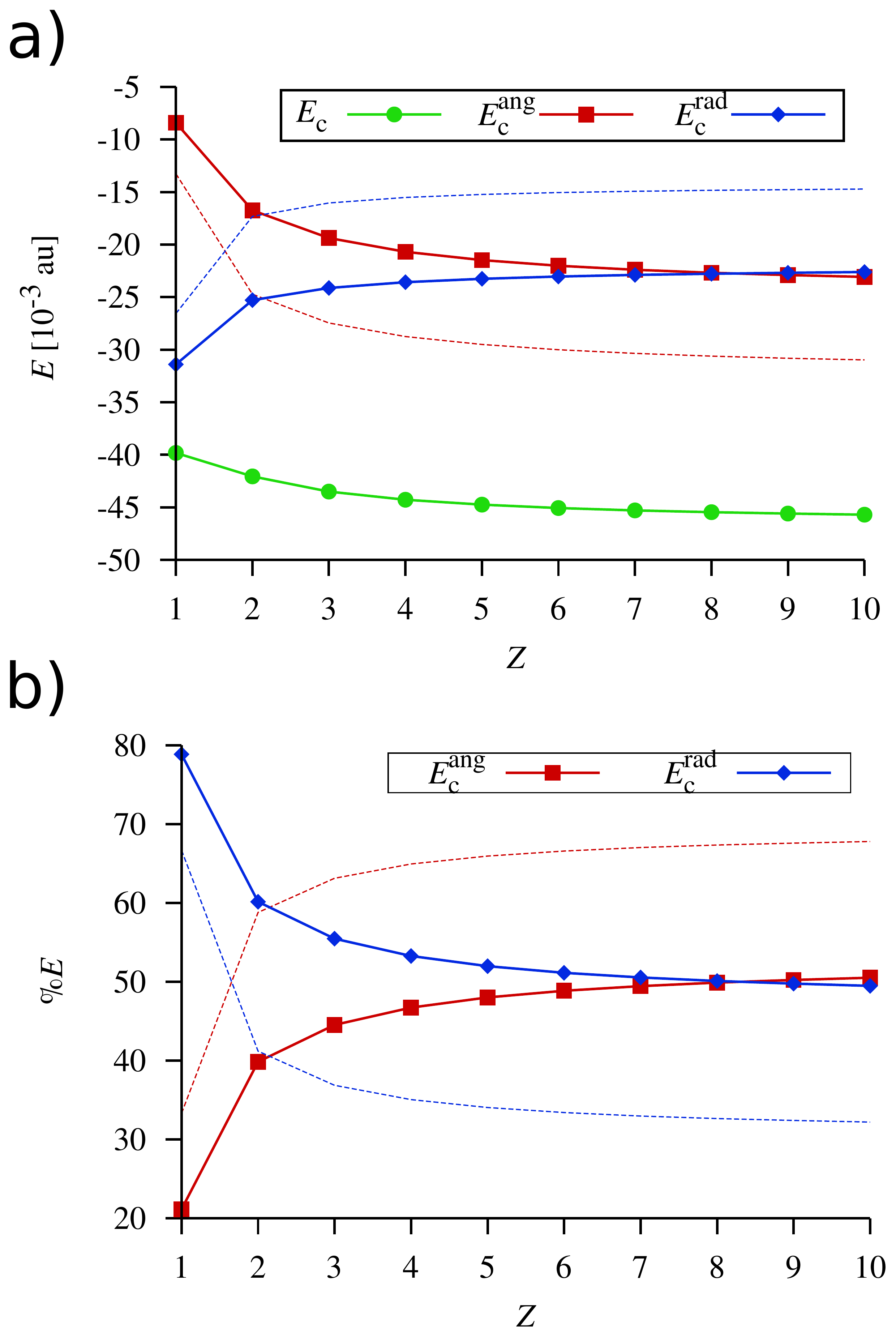}
\caption{Variation of correlation energies across two-electron atoms: a)
Solid red line corresponds to  $E_c^{\rm ang}$ estimated using the 
virial theorem as $-T^{\rm ang}$ while the dotted red line corresponds to the 
deviation of the energy computed using $\Psi^{\rm rad}$
from the exact energy. Solid blue line points to $E_c^{\rm rad}$  computed as  
$E_c-E_c^{\rm ang}$, while the blue dotted line is the deviation of the energy computed 
using $\Psi^{\rm rad}$  from the HF energy; b) contributions to $E_c$ are shown in percentages.
}
\label{fig:Ec}
\end{figure}

Scanning through the two-electron atoms H$^{-}$ until Ne$^{8+}$, we have computed $E_c^{\rm rad}$ and $E_c^{\rm ang}$, using both: the conventional approach wherein the contribution from the radial-angular coupling is associated with the angular term, and using the new formalism proposed in this study that is free of such coupling. The results are displayed in Fig.~\ref{fig:Ec}. As the most striking feature of this figure, one notes the conventional estimation of $E_c^{\rm ang}$ to be more negative than the exact result; while to the same extent, conventional estimation of $E_c^{\rm rad}$ less negative than the exact one.  This trend can be understood as follows: Briefly, for  helium, exact separation of the kinetic energy operator results in $E_c^{\rm rad}=0.602E_c$ and $E_c^{\rm ang}=0.398E_c$.  In contrary, previous conventions suggest  $E_c^{\rm rad}=0.413E_c$ and $E_c^{\rm ang}=0.587E_c$ undermining the importance of radial interactions over the angular one.  Our analysis reveals the conventional $E_c^{\rm ang}$ to include $0.378 E_c$ arising from radial-angular coupling, and half of this value must be added to the conventional $E_c^{\rm rad}$ to predict the exact value correctly.

Furthermore, we find the total correlation energy to increase with $Z$ for lighter atoms, but converging already near $Z=10$ (see Fig.~\ref{fig:Ec}).  The same plot also reveals  the individual radial and angular components to also converge, with deviations of less than  0.0001 au between F$^{7+}$ and Ne$^{8+}$. 
Owing to a somewhat unbounded nature, we find the radial correlation to be dominant for the lightest system, H$^-$, with $E_c^{\rm rad}$ accounting for 78.9\% of $E_c$. 



In conclusion, we present a new strategy to partition the correlation energy of two-electron atoms into radial and angular contributions. We have shown previous estimations of radial correlation energy of helium, based on a  limiting radial wavefunction, to underestimate the exact value due to the neglect of radial-angular coupling thereby suggesting $E_c^{\rm ang}$ to be larger in magnitude than $E_c^{\rm rad}$. Since an HF wavefunction is entirely devoid of the angular interaction, the corresponding kinetic energy of two-electron atoms arises exclusively from many-body correlation. In fact, this term is one of the essential ingredients of the hitherto unknown exact exchange-correlation (XC) functional in the density functional theory~\cite{parr1989density}.  It will be of interest to see if the presented results aid in the design of modern XC functionals predicting correct angular kinetic energy, at least for the limiting case of two-electron atoms.  

ARK gratefully acknowledges a summer fellowship of TIFR Visiting Students' Research Programme (VSRP). RR thanks TIFR for financial support. 

\bibliography{lit} 

\begin{thebibliography}{23}%
\makeatletter
\providecommand \@ifxundefined [1]{%
 \@ifx{#1\undefined}
}%
\providecommand \@ifnum [1]{%
 \ifnum #1\expandafter \@firstoftwo
 \else \expandafter \@secondoftwo
 \fi
}%
\providecommand \@ifx [1]{%
 \ifx #1\expandafter \@firstoftwo
 \else \expandafter \@secondoftwo
 \fi
}%
\providecommand \natexlab [1]{#1}%
\providecommand \enquote  [1]{``#1''}%
\providecommand \bibnamefont  [1]{#1}%
\providecommand \bibfnamefont [1]{#1}%
\providecommand \citenamefont [1]{#1}%
\providecommand \href@noop [0]{\@secondoftwo}%
\providecommand \href [0]{\begingroup \@sanitize@url \@href}%
\providecommand \@href[1]{\@@startlink{#1}\@@href}%
\providecommand \@@href[1]{\endgroup#1\@@endlink}%
\providecommand \@sanitize@url [0]{\catcode `\\12\catcode `\$12\catcode
  `\&12\catcode `\#12\catcode `\^12\catcode `\_12\catcode `\%12\relax}%
\providecommand \@@startlink[1]{}%
\providecommand \@@endlink[0]{}%
\providecommand \url  [0]{\begingroup\@sanitize@url \@url }%
\providecommand \@url [1]{\endgroup\@href {#1}{\urlprefix }}%
\providecommand \urlprefix  [0]{URL }%
\providecommand \Eprint [0]{\href }%
\providecommand \doibase [0]{http://dx.doi.org/}%
\providecommand \selectlanguage [0]{\@gobble}%
\providecommand \bibinfo  [0]{\@secondoftwo}%
\providecommand \bibfield  [0]{\@secondoftwo}%
\providecommand \translation [1]{[#1]}%
\providecommand \BibitemOpen [0]{}%
\providecommand \bibitemStop [0]{}%
\providecommand \bibitemNoStop [0]{.\EOS\space}%
\providecommand \EOS [0]{\spacefactor3000\relax}%
\providecommand \BibitemShut  [1]{\csname bibitem#1\endcsname}%
\let\auto@bib@innerbib\@empty
\bibitem [{\citenamefont {Yan}\ and\ \citenamefont
  {Drake}(1995)}]{yan1995high}%
  \BibitemOpen
  \bibfield  {author} {\bibinfo {author} {\bibfnamefont {Z.-C.}\ \bibnamefont
  {Yan}}\ and\ \bibinfo {author} {\bibfnamefont {G.}~\bibnamefont {Drake}},\
  }\href@noop {} {\bibfield  {journal} {\bibinfo  {journal} {Phys. Rev. Lett.}\
  }\textbf {\bibinfo {volume} {74}},\ \bibinfo {pages} {4791} (\bibinfo {year}
  {1995})}\BibitemShut {NoStop}%
\bibitem [{\citenamefont {Hylleraas}(1964)}]{hylleraas1964schrodinger}%
  \BibitemOpen
  \bibfield  {author} {\bibinfo {author} {\bibfnamefont {E.~A.}\ \bibnamefont
  {Hylleraas}},\ }\href@noop {} {\bibfield  {journal} {\bibinfo  {journal}
  {Adv. Quantum Chem.}\ }\textbf {\bibinfo {volume} {1}},\ \bibinfo {pages} {1}
  (\bibinfo {year} {1964})}\BibitemShut {NoStop}%
\bibitem [{\citenamefont {Nakashima}\ and\ \citenamefont
  {Nakatsuji}(2007)}]{nakashima2007solving}%
  \BibitemOpen
  \bibfield  {author} {\bibinfo {author} {\bibfnamefont {H.}~\bibnamefont
  {Nakashima}}\ and\ \bibinfo {author} {\bibfnamefont {H.}~\bibnamefont
  {Nakatsuji}},\ }\href@noop {} {\bibfield  {journal} {\bibinfo  {journal} {J.
  Chem. Phys.}\ }\textbf {\bibinfo {volume} {127}},\ \bibinfo {pages} {224104}
  (\bibinfo {year} {2007})}\BibitemShut {NoStop}%
\bibitem [{\citenamefont {Raffenetti}(1973)}]{raffenetti1973even}%
  \BibitemOpen
  \bibfield  {author} {\bibinfo {author} {\bibfnamefont {R.~C.}\ \bibnamefont
  {Raffenetti}},\ }\href@noop {} {\bibfield  {journal} {\bibinfo  {journal} {J.
  Chem. Phys.}\ }\textbf {\bibinfo {volume} {59}},\ \bibinfo {pages} {5936}
  (\bibinfo {year} {1973})}\BibitemShut {NoStop}%
\bibitem [{\citenamefont {Taylor}\ and\ \citenamefont
  {Parr}(1952)}]{taylor1952superposition}%
  \BibitemOpen
  \bibfield  {author} {\bibinfo {author} {\bibfnamefont {G.~R.}\ \bibnamefont
  {Taylor}}\ and\ \bibinfo {author} {\bibfnamefont {R.~G.}\ \bibnamefont
  {Parr}},\ }\href@noop {} {\bibfield  {journal} {\bibinfo  {journal} {Proc.
  Natl. Acad. Sci. USA}\ }\textbf {\bibinfo {volume} {38}},\ \bibinfo {pages}
  {154} (\bibinfo {year} {1952})}\BibitemShut {NoStop}%
\bibitem [{\citenamefont {Goldman}(1997)}]{goldman1997accurate}%
  \BibitemOpen
  \bibfield  {author} {\bibinfo {author} {\bibfnamefont {S.}~\bibnamefont
  {Goldman}},\ }\href@noop {} {\bibfield  {journal} {\bibinfo  {journal} {Phys.
  Rev. Lett.}\ }\textbf {\bibinfo {volume} {78}},\ \bibinfo {pages} {2325}
  (\bibinfo {year} {1997})}\BibitemShut {NoStop}%
\bibitem [{\citenamefont {Lennard-Jones}\ and\ \citenamefont
  {Pople}(1952)}]{lennard1952liii}%
  \BibitemOpen
  \bibfield  {author} {\bibinfo {author} {\bibfnamefont {J.}~\bibnamefont
  {Lennard-Jones}}\ and\ \bibinfo {author} {\bibfnamefont {J.}~\bibnamefont
  {Pople}},\ }\href@noop {} {\bibfield  {journal} {\bibinfo  {journal} {Phil.
  Mag.}\ }\textbf {\bibinfo {volume} {43}},\ \bibinfo {pages} {581} (\bibinfo
  {year} {1952})}\BibitemShut {NoStop}%
\bibitem [{\citenamefont {Moiseyev}\ and\ \citenamefont
  {Katriel}(1975)}]{moiseyev1975coupling}%
  \BibitemOpen
  \bibfield  {author} {\bibinfo {author} {\bibfnamefont {N.}~\bibnamefont
  {Moiseyev}}\ and\ \bibinfo {author} {\bibfnamefont {J.}~\bibnamefont
  {Katriel}},\ }\href@noop {} {\bibfield  {journal} {\bibinfo  {journal} {Chem.
  Phys.}\ }\textbf {\bibinfo {volume} {10}},\ \bibinfo {pages} {67} (\bibinfo
  {year} {1975})}\BibitemShut {NoStop}%
\bibitem [{\citenamefont {Wilson}\ \emph {et~al.}(2010)\citenamefont {Wilson},
  \citenamefont {Montgomery}, \citenamefont {Sen},\ and\ \citenamefont
  {Thompson}}]{wilson2010electron}%
  \BibitemOpen
  \bibfield  {author} {\bibinfo {author} {\bibfnamefont {C.}~\bibnamefont
  {Wilson}}, \bibinfo {author} {\bibfnamefont {H.}~\bibnamefont {Montgomery}},
  \bibinfo {author} {\bibfnamefont {K.}~\bibnamefont {Sen}}, \ and\ \bibinfo
  {author} {\bibfnamefont {D.}~\bibnamefont {Thompson}},\ }\href@noop {}
  {\bibfield  {journal} {\bibinfo  {journal} {Phys. Lett.}\ }\textbf {\bibinfo
  {volume} {374}},\ \bibinfo {pages} {4415} (\bibinfo {year}
  {2010})}\BibitemShut {NoStop}%
\bibitem [{\citenamefont {Saha}\ \emph {et~al.}(2003)\citenamefont {Saha},
  \citenamefont {Bhattacharyya}, \citenamefont {Mukherjee},\ and\ \citenamefont
  {Mukherjee}}]{saha2003radial}%
  \BibitemOpen
  \bibfield  {author} {\bibinfo {author} {\bibfnamefont {B.}~\bibnamefont
  {Saha}}, \bibinfo {author} {\bibfnamefont {S.}~\bibnamefont {Bhattacharyya}},
  \bibinfo {author} {\bibfnamefont {T.}~\bibnamefont {Mukherjee}}, \ and\
  \bibinfo {author} {\bibfnamefont {P.}~\bibnamefont {Mukherjee}},\ }\href@noop
  {} {\bibfield  {journal} {\bibinfo  {journal} {Int. J. Quantum Chem.}\
  }\textbf {\bibinfo {volume} {92}},\ \bibinfo {pages} {413} (\bibinfo {year}
  {2003})}\BibitemShut {NoStop}%
\bibitem [{\citenamefont {Green}\ \emph {et~al.}(1953)\citenamefont {Green},
  \citenamefont {Mulder},\ and\ \citenamefont {Milner}}]{green1953correlation}%
  \BibitemOpen
  \bibfield  {author} {\bibinfo {author} {\bibfnamefont {L.~C.}\ \bibnamefont
  {Green}}, \bibinfo {author} {\bibfnamefont {M.~M.}\ \bibnamefont {Mulder}}, \
  and\ \bibinfo {author} {\bibfnamefont {P.~C.}\ \bibnamefont {Milner}},\
  }\href@noop {} {\bibfield  {journal} {\bibinfo  {journal} {Phys. Rev.}\
  }\textbf {\bibinfo {volume} {91}},\ \bibinfo {pages} {35} (\bibinfo {year}
  {1953})}\BibitemShut {NoStop}%
\bibitem [{\citenamefont {Green}\ \emph {et~al.}(1954)\citenamefont {Green},
  \citenamefont {Lewis}, \citenamefont {Mulder}, \citenamefont {Wyeth},\ and\
  \citenamefont {Woll~Jr}}]{green1954correlation}%
  \BibitemOpen
  \bibfield  {author} {\bibinfo {author} {\bibfnamefont {L.~C.}\ \bibnamefont
  {Green}}, \bibinfo {author} {\bibfnamefont {M.~N.}\ \bibnamefont {Lewis}},
  \bibinfo {author} {\bibfnamefont {M.~M.}\ \bibnamefont {Mulder}}, \bibinfo
  {author} {\bibfnamefont {C.~W.}\ \bibnamefont {Wyeth}}, \ and\ \bibinfo
  {author} {\bibfnamefont {J.~W.}\ \bibnamefont {Woll~Jr}},\ }\href@noop {}
  {\bibfield  {journal} {\bibinfo  {journal} {Phys. Rev.}\ }\textbf {\bibinfo
  {volume} {93}},\ \bibinfo {pages} {273} (\bibinfo {year} {1954})}\BibitemShut
  {NoStop}%
\bibitem [{\citenamefont {Chandrasekhar}\ \emph {et~al.}(1953)\citenamefont
  {Chandrasekhar}, \citenamefont {Elbert},\ and\ \citenamefont
  {Herzberg}}]{chandrasekhar1953shift}%
  \BibitemOpen
  \bibfield  {author} {\bibinfo {author} {\bibfnamefont {S.}~\bibnamefont
  {Chandrasekhar}}, \bibinfo {author} {\bibfnamefont {D.}~\bibnamefont
  {Elbert}}, \ and\ \bibinfo {author} {\bibfnamefont {G.}~\bibnamefont
  {Herzberg}},\ }\href@noop {} {\bibfield  {journal} {\bibinfo  {journal}
  {Physical Review}\ }\textbf {\bibinfo {volume} {91}},\ \bibinfo {pages}
  {1172} (\bibinfo {year} {1953})}\BibitemShut {NoStop}%
\bibitem [{\citenamefont {Drake}\ \emph {et~al.}(2002)\citenamefont {Drake},
  \citenamefont {Cassar},\ and\ \citenamefont {Nistor}}]{drake2002ground}%
  \BibitemOpen
  \bibfield  {author} {\bibinfo {author} {\bibfnamefont {G.~W.}\ \bibnamefont
  {Drake}}, \bibinfo {author} {\bibfnamefont {M.~M.}\ \bibnamefont {Cassar}}, \
  and\ \bibinfo {author} {\bibfnamefont {R.~A.}\ \bibnamefont {Nistor}},\
  }\href@noop {} {\bibfield  {journal} {\bibinfo  {journal} {Physical Review
  A}\ }\textbf {\bibinfo {volume} {65}},\ \bibinfo {pages} {054501} (\bibinfo
  {year} {2002})}\BibitemShut {NoStop}%
\bibitem [{\citenamefont {Kinoshita}(1957)}]{kinoshita1957ground}%
  \BibitemOpen
  \bibfield  {author} {\bibinfo {author} {\bibfnamefont {T.}~\bibnamefont
  {Kinoshita}},\ }\href@noop {} {\bibfield  {journal} {\bibinfo  {journal}
  {Phys. Rev.}\ }\textbf {\bibinfo {volume} {105}},\ \bibinfo {pages} {1490}
  (\bibinfo {year} {1957})}\BibitemShut {NoStop}%
\bibitem [{\citenamefont {Drake}\ and\ \citenamefont
  {Van}(1994)}]{drake1994variational}%
  \BibitemOpen
  \bibfield  {author} {\bibinfo {author} {\bibfnamefont {G.}~\bibnamefont
  {Drake}}\ and\ \bibinfo {author} {\bibfnamefont {Z.-C.}\ \bibnamefont
  {Van}},\ }\href@noop {} {\bibfield  {journal} {\bibinfo  {journal} {Chem.
  Phys. Lett.}\ }\textbf {\bibinfo {volume} {229}},\ \bibinfo {pages} {486}
  (\bibinfo {year} {1994})}\BibitemShut {NoStop}%
\bibitem [{\citenamefont {Koga}(1996)}]{koga1996radial}%
  \BibitemOpen
  \bibfield  {author} {\bibinfo {author} {\bibfnamefont {T.}~\bibnamefont
  {Koga}},\ }\href@noop {} {\bibfield  {journal} {\bibinfo  {journal} {Z. f.
  Phys.}\ }\textbf {\bibinfo {volume} {37}},\ \bibinfo {pages} {301} (\bibinfo
  {year} {1996})}\BibitemShut {NoStop}%
\bibitem [{\citenamefont {Hahn}(2005)}]{hahn2005cuba}%
  \BibitemOpen
  \bibfield  {author} {\bibinfo {author} {\bibfnamefont {T.}~\bibnamefont
  {Hahn}},\ }\href@noop {} {\bibfield  {journal} {\bibinfo  {journal} {Comput.
  Phys. Commun.}\ }\textbf {\bibinfo {volume} {168}},\ \bibinfo {pages} {78}
  (\bibinfo {year} {2005})}\BibitemShut {NoStop}%
\bibitem [{\citenamefont {Bethe}\ and\ \citenamefont
  {Salpeter}(2008)}]{bethe2012quantum}%
  \BibitemOpen
  \bibfield  {author} {\bibinfo {author} {\bibfnamefont {H.~A.}\ \bibnamefont
  {Bethe}}\ and\ \bibinfo {author} {\bibfnamefont {E.~E.}\ \bibnamefont
  {Salpeter}},\ }\href@noop {} {\emph {\bibinfo {title} {Quantum mechanics of
  one-and two-electron atoms}}}\ (\bibinfo  {publisher} {Dover},\ \bibinfo
  {year} {2008})\BibitemShut {NoStop}%
\bibitem [{\citenamefont {Koga}(2010)}]{koga2010electron}%
  \BibitemOpen
  \bibfield  {author} {\bibinfo {author} {\bibfnamefont {T.}~\bibnamefont
  {Koga}},\ }\href@noop {} {\bibfield  {journal} {\bibinfo  {journal} {Journal
  of Molecular Structure: THEOCHEM}\ }\textbf {\bibinfo {volume} {947}},\
  \bibinfo {pages} {115} (\bibinfo {year} {2010})}\BibitemShut {NoStop}%
\bibitem [{\citenamefont {Koga}\ \emph {et~al.}(2011)\citenamefont {Koga},
  \citenamefont {Matsuyama},\ and\ \citenamefont
  {Thakkar}}]{koga2011interelectronic}%
  \BibitemOpen
  \bibfield  {author} {\bibinfo {author} {\bibfnamefont {T.}~\bibnamefont
  {Koga}}, \bibinfo {author} {\bibfnamefont {H.}~\bibnamefont {Matsuyama}}, \
  and\ \bibinfo {author} {\bibfnamefont {A.~J.}\ \bibnamefont {Thakkar}},\
  }\href@noop {} {\bibfield  {journal} {\bibinfo  {journal} {Chem. Phys.
  Lett.}\ }\textbf {\bibinfo {volume} {512}},\ \bibinfo {pages} {287} (\bibinfo
  {year} {2011})}\BibitemShut {NoStop}%
\bibitem [{\citenamefont {Ramakrishnan}\ and\ \citenamefont
  {Nest}(2012)}]{ramakrishnan2012control}%
  \BibitemOpen
  \bibfield  {author} {\bibinfo {author} {\bibfnamefont {R.}~\bibnamefont
  {Ramakrishnan}}\ and\ \bibinfo {author} {\bibfnamefont {M.}~\bibnamefont
  {Nest}},\ }\href@noop {} {\bibfield  {journal} {\bibinfo  {journal} {Phys.
  Rev. A}\ }\textbf {\bibinfo {volume} {85}},\ \bibinfo {pages} {054501}
  (\bibinfo {year} {2012})}\BibitemShut {NoStop}%
\bibitem [{\citenamefont {Parr}\ and\ \citenamefont
  {Weitao}(1989)}]{parr1989density}%
  \BibitemOpen
  \bibfield  {author} {\bibinfo {author} {\bibfnamefont {R.~G.}\ \bibnamefont
  {Parr}}\ and\ \bibinfo {author} {\bibfnamefont {Y.}~\bibnamefont {Weitao}},\
  }\href@noop {} {\emph {\bibinfo {title} {Density-Functional Theory of Atoms
  and Molecules}}}\ (\bibinfo  {publisher} {Oxford University Press},\ \bibinfo
  {year} {1989})\BibitemShut {NoStop}%
\end{thebibliography}%

\end{document}